\documentclass[aps, prb, 10pt, preprint, preprintnumbers,
superscriptaddress, amsmath, amssymb, letterpaper]{revtex4-1}
\usepackage[pdftex]{graphicx}
\usepackage[ngerman,english]{babel}
\usepackage[ansinew]{inputenc}
\usepackage{natbib}
\usepackage{color}
\usepackage{siunitx}
\usepackage{calc}
\usepackage[colorlinks=true, pdfstartview=FitV, linkcolor=blue,
citecolor=blue, urlcolor=blue]{hyperref}

\definecolor{green2}{rgb}{.0, .58, 0}

\begin{document}

\title{Observation of vortex-nucleated magnetization reversal in
  individual ferromagnetic nanotubes}

\author{A.~Mehlin} \thanks{Authors contributed equally.}
\affiliation{Department of Physics, University of Basel, 4056 Basel,
  Switzerland}
		
\author{B.~Gross} \thanks{Authors contributed equally.}
\affiliation{Department of Physics, University of Basel, 4056 Basel,
  Switzerland}

\author{M.~Wyss} \affiliation{Department of Physics, University of
  Basel, 4056 Basel, Switzerland}
  
\author{T.~Schefer} \affiliation{Department of Physics, University of
  Basel, 4056 Basel, Switzerland}
  
\author{G.~T\"{u}t\"{u}nc\"{u}oglu} \affiliation{Laboratory of
  Semiconductor Materials, Institute of Materials (IMX), School of
  Engineering, \'Ecole Polytechnique F\'ed\'erale de Lausanne (EPFL),
  1015 Lausanne, Switzerland}
  
\author{F.~Heimbach} \affiliation{Lehrstuhl f\"{u}r Physik
  funktionaler Schichtsysteme, Physik Department E10, Technische
  Universit\"{a}t M\"{u}nchen, 85747 Garching, Germany}
	
\author{A.~Fontcuberta~i~Morral} \affiliation{Laboratory of
  Semiconductor Materials, Institute of Materials (IMX), School of
  Engineering, \'Ecole Polytechnique F\'ed\'erale de Lausanne (EPFL),
  1015 Lausanne, Switzerland}
	
\author{D.~Grundler} \affiliation{Laboratory of Nanoscale Magnetic
  Materials and Magnonics, Institute of Materials (IMX) and Institute
  of Microengineering (IMT), School of Engineering, \'Ecole
  Polytechnique F\'ed\'erale de Lausanne (EPFL), 1015 Lausanne,
  Switzerland}
	
\author{M.~Poggio} \affiliation{Department of Physics, University of
  Basel, 4056 Basel, Switzerland} \email{martino.poggio@unibas.ch}
\homepage{http://poggiolab.unibas.ch/}

\begin{abstract} 
  The reversal of a uniform axial magnetization in a ferromagnetic
  nanotube (FNT) has been predicted to nucleate and propagate through
  vortex domains forming at the ends.  In dynamic cantilever
  magnetometry measurements of individual FNTs, we identify the entry
  of these vortices as a function of applied magnetic field and show
  that they mark the nucleation of magnetization reversal.  We find
  that the entry field depends sensitively on the angle between the
  end surface of the FNT and the applied field.  Micromagnetic
  simulations substantiate the experimental results and highlight the
  importance of the ends in determining the reversal process.  The
  control over end vortex formation enabled by our findings is
  promising for the production of FNTs with tailored reversal
  properties.
\end{abstract} 

\maketitle

The study of magnetization reversal in magnetic nanostructures is a
topic of major fundamental and practical interest.  In particular, a
controllable, fast, and reproducible reversal is crucial for
applications in high density magnetic storage.  This process, however,
is often conditioned by the presence of edge and surface domains.
Near borders, magnetization tends to change direction in order to
minimize stray field energy.  As a result, the form of surfaces and
edges -- including any imperfections or roughness -- determines the
configuration of the magnetization in their vicinity.  The resulting
magnetization inhomogeneities tend to affect reversal by acting as
nucleation sites for complex switching
processes~\cite{coey_magnetism_2010,rothman_observation_2001}.
Furthermore, small differences in the initial configurations of edge
and surface domains can lead to entirely different reversal modes,
complicating the control and reproducibility of magnetic switching
from nanomagnet to nanomagnet~\cite{zheng_switching_1997}.

The high surface-to-volume ratio of magnetic nanostructures makes
mitigating these effects essential in the design of high-density
memory elements.  One way to reduce the effect of edges and surfaces
on magnetic reversal is to use magnetic structures that support
flux-closure magnetization configurations~\cite{kent_annular_2011}.
Since these configurations minimize stray field, edges and surfaces
play a minor role in determining both their equilibrium state and
their dynamics.  Ferromagnetic nanotubes (FNTs) are one type of
nanostructure supporting such states.  In particular, reversal of
uniform axial configurations in FNTs has been predicted to nucleate
and propagate through vortex configurations, which appear at the FNT
ends and whose magnetization curls around their hollow
core~\cite{usov_domain_2007,landeros_reversal_2007,landeros_equilibrium_2009,landeros_domain_2010}.
Theory has so far only considered FNTs with perfect, flat ends,
despite their importance as the nucleation sites of the reversal.
Here, we show the experimental signatures of this nucleation and
reveal its dependence on the angle of the FNT ends.  Magnetization
reversal in FNTs offers some potential advantages over the equivalent
and well-understood process in ferromagnetic nanowires: in particular,
the core-free geometry of FNTs has been predicted to favor uniform
switching fields and high
reproducibility~\cite{wang_spin_2005,escrig_phase_2007,landeros_reversal_2007}.
Understanding and controlling the switching process in real FNTs is a
crucial step in enabling practical applications.

We study magnetization reversal in individual FNTs using dynamic
cantilever magnetometry (DCM).  This technique involves a measurement
of the mechanical resonance frequency $f$ of a cantilever, to which
the FNT of interest has been attached, as a function of a uniform
externally applied magnetic field $\mathbf{H}$.  The frequency shift
$\Delta f = f - f_0$, where $f_0$ is the resonance frequency at $H =
0$, reveals the curvature of the magnetic energy with respect to
rotations about the cantilever oscillation
axis~\cite{mehlin_stabilized_2015,gross_dynamic_2016}:
\begin{equation}
  \Delta f = \frac{f_0}{2 k_0 l_e^2} \left ( \left.  \frac{\partial^2
        E_m}{\partial \theta_c^2} \right|_{\theta_c=0} \right ),
\label{DCM_equation}
\end{equation}
where $k_0$ is the cantilever's spring constant, $l_e$ its effective
length, and $\theta_c$ its angle of oscillation.  We simulate the DCM
measurements by constructing a micromagnetic model of the experiment
with the software package
\textit{Mumax3}~\cite{vansteenkiste_design_2014}, which employs the
Landau-Lifshitz-Gilbert micromagnetic formalism using
finite-difference discretization.  The simulations allow us to relate
DCM signal to the magnetization configurations present in a FNT.
These insights -- combined with the high torque sensitivity provided
by ultrasoft Si cantilevers -- uncover magnetization reversal behavior
of an individual FNT.
\begin{figure}
\includegraphics{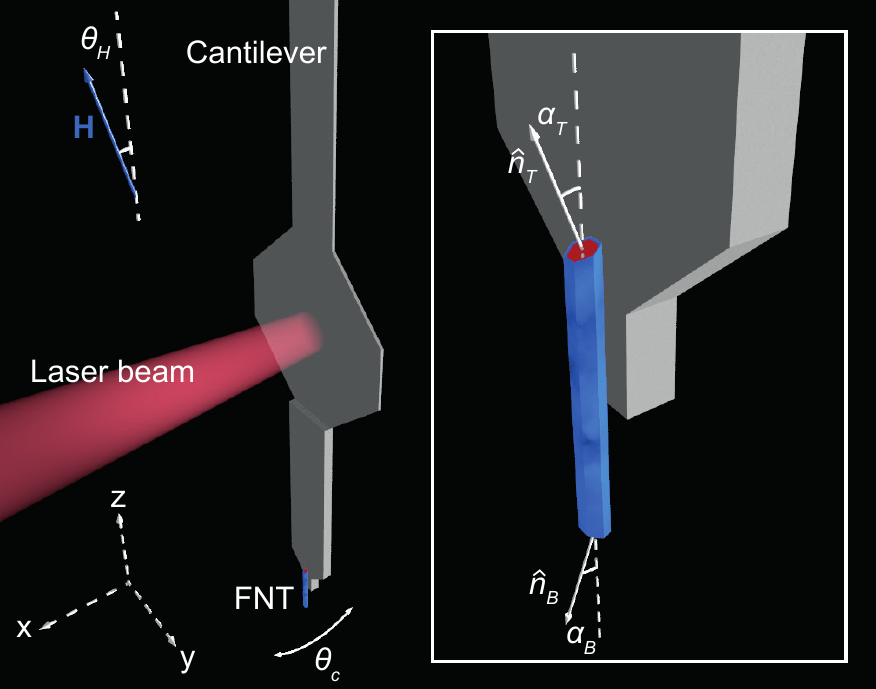}
\caption{Schematic diagram of the measurement setup: Si cantilever
  (gray) and CoFeB FNT (blue) with GaAs core (red).  The cantilever
  oscillates about $\hat{y}$ and the FNT axis is parallel to
  $\hat{z}$.  The externally applied magnetic field $\mathbf{H}$ can
  be rotated in the $xz$-plane by an angle $\theta_H$ with respect to
  $\hat{z}$.  The top (bottom) end of the FNT lies in a plane
  perpendicular to $\hat{n}_T$ ($\hat{n}_B$).}
\label{fig:setup}
\end{figure}

FNT samples consist of a 30-nm-thick ferromagnetic shell of CoFeB
surrounding a non-magnetic GaAs core with hexagonal cross-section.
The amorphous and homogeneous CoFeB shell is magnetron sputtered onto
template GaAs nanowires, which are grown by molecular beam
epitaxy~\cite{ruffer_anisotropic_2014}.  Scanning electron micrographs
(SEMs) of the studied FNTs reveal continuous and defect-free surfaces,
whose roughness is less than 5~nm~\cite{SuppMat}.  The FNTs have a
diameter, which we define as the diameter of the circle circumscribing
their hexagonal cross-section, between 270 and 300 nm.  Lengths from
0.6 to 2.9~\si{\micro\meter} are obtained by cutting individual FNTs into
segments using a focused ion beam (FIB)~\cite{wyss_imaging_2017}.
This procedure ensures FNTs with smooth and well-defined ends, which
-- in general -- are tilted relative to the plane normal to the FNT
axis, as shown in Fig.~\ref{fig:setup}.  After cutting, each FNT is
affixed to the end of an ultrasoft Si cantilever, which is mounted in
the DCM measurement setup.
\begin{figure}
  \includegraphics[]{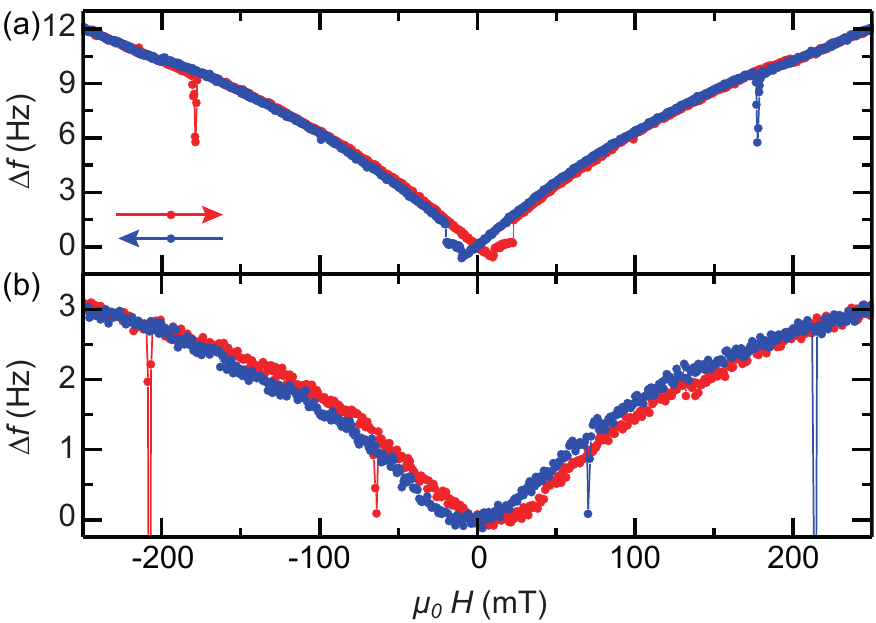}
  \caption{Magnetic reversal of (a) a 2.2-\si{\micro\meter}-long and (b) a
    0.6-\si{\micro\meter}-long FNT measured by DCM at 280~K.  $\mathbf{H}$ is
    applied approximately along $\hat{z}$.  As in all following
    figures, color-coded arrows denote the direction that the magnetic
    field is stepped.}
  \label{fig:DCM_exp}
\end{figure}

Fig.~\ref{fig:DCM_exp} shows DCM measurements at 280~K of two FNTs of
different lengths: (a) 2.2~\si{\micro\meter} and (b)
0.6~\si{\micro\meter}.  For each FNT, measured $\Delta f (H)$ is
plotted for $\mathbf{H}$ applied approximately along its long axis
$\hat{z}$ and swept in the positive and negative direction.  Since the
cantilevers used here have similar mechanical properties, the
magnitude of the frequency response is roughly proportional to the FNT
length and therefore to the volume of magnetic material.  Three major
characteristics can be identified in the data sets. First, both show
an overall V-shape, consistent with the near coincidence of the FNT
easy axis and $\mathbf{H}$~\cite{gross_dynamic_2016}.  Second, one or
two spikes toward negative $\Delta f$ occur in forward applied field
between $\pm 220$ and $\pm 60$~mT as well as weak echos of these
features in reverse magnetic field.  Third, around zero field, where
the slope of $\Delta f (H)$ inverts, a distinct difference between the
two FNTs is evident.  The shorter FNT shows a parabolic dependence,
without ever becoming negative, while the longer one crosses to
negative values of $\Delta f$ before exhibiting two discontinuous
steps.  The latter behavior is similar to that found for an even
longer 2.9-\si{\micro\meter}-long FNT~\cite{SuppMat}.  Measurements on
FNTs of all three lengths were carried out at 4~K with similar
results~\cite{SuppMat}.

\begin{figure*}
  \includegraphics[]{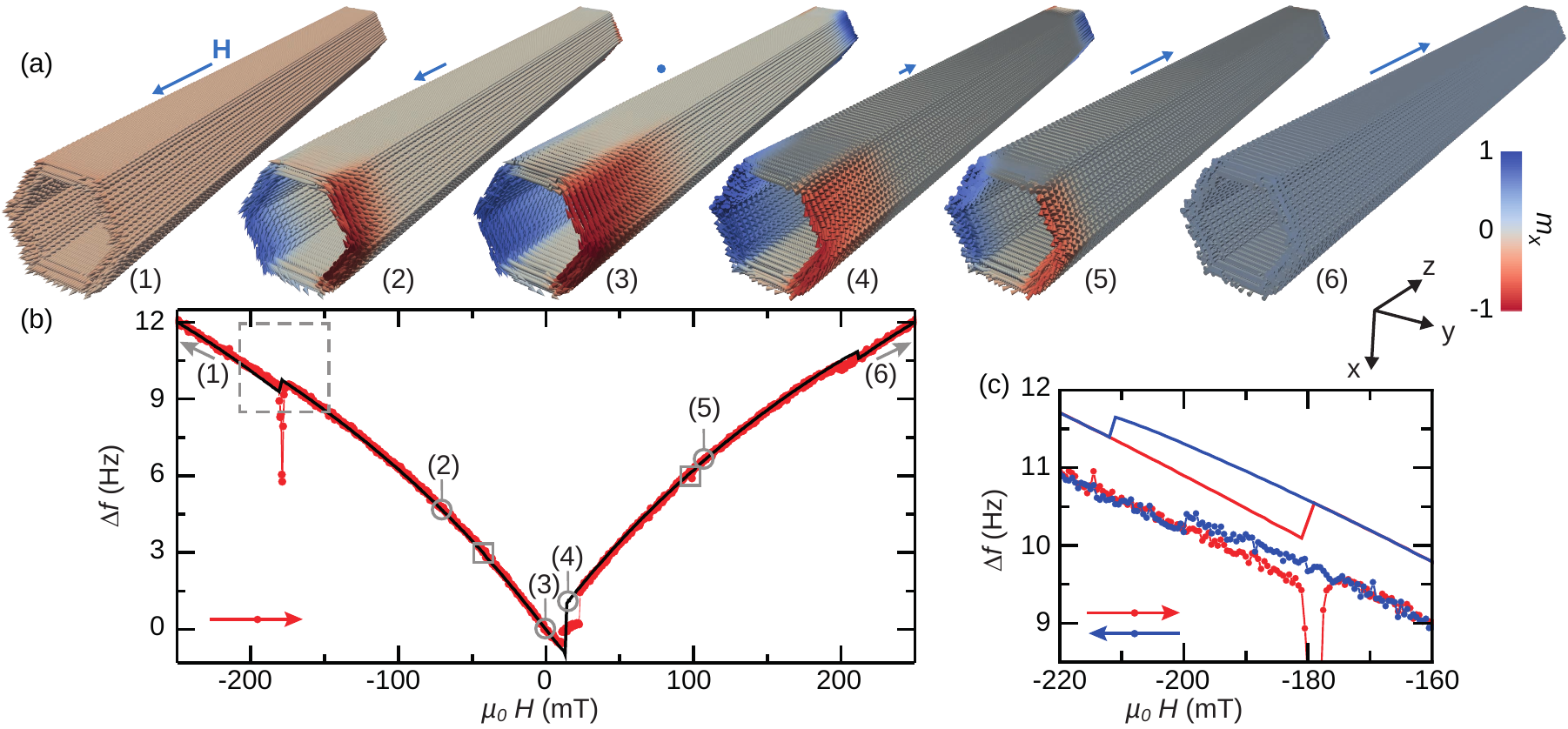}
  \caption{Simulated and measured reversal of a 2.2-\si{\micro\meter}-long FNT.
    (a) Calculated magnetization configurations for $\Delta f(H)$
    corresponding to the labels.  (b) Simulated (lines) and measured
    (points) DCM signal.  Squares highlight simulated vortex
    entry/exit features, which are difficult to see.  (c) A detailed
    view of DCM signatures of vortex entrance and exit.  For the simulation, $\alpha_T =
    6.5^\circ$, $\alpha_B = 10.5^\circ$, and $\theta_H = 11.0^\circ$.}
\label{fig:2um_DCM_exp}
\end{figure*}
The dimensions of a FNT are predicted to have a determining influence
on its magnetic reversal.  In particular, FNTs with a larger than
critical diameter, reverse via nucleation of vortex rather than
transverse domain
walls~\cite{landeros_equilibrium_2009,landeros_domain_2010}.  Since
this diameter ranges from a few nanometers to 20 nm, all
experimentally fabricated FNTs should reverse through vortex domains.
For long FNTs, i.e.\ 2~\si{\micro\meter} or longer for our
cross-sectional geometry, the expected progression of the
magnetization for $\mathbf{H}$ approximately along $\hat{z}$ can be
summarized as shown in Fig.~\ref{fig:2um_DCM_exp}~(a).  This specific
progression is the result of our simulations, but similar progressions
were predicted by previous analytical and numerical
models~\cite{landeros_equilibrium_2009,gross_dynamic_2016}.  Starting
from full saturation at negative $H$, vortices enter the two tube
ends, setting the nucleation field of the reversal.  At these fields,
both relative circulation senses of the end domains have equal energy,
such that the appearance of one or the other is likely driven by
sample imperfections.  As $H$ approaches zero and becomes positive,
the vortices grow along the tube axis toward the center.  At a small
positive reverse field, the magnetization in the central axial domain
irreversibly inverts, while the end vortices persist.  From here on,
the vortices shrink in size with progressively larger positive $H$,
until they exit the FNT, marking the end of the reversal.

Fig.~\ref{fig:2um_DCM_exp}~(b) shows measured and simulated $\Delta f
(H)$ for the 2.2-\si{\micro\meter}-long FNT as $H$ is swept in the positive
direction.  The simulated $\Delta f (H)$ is calculated using the
measured properties of the cantilever and the geometrical and material
parameters of the FNT (adjusted within their error).  Numerical labels
indicate the magnetization configuration in
Fig.~\ref{fig:2um_DCM_exp}~(a) corresponding to a particular value of
$H$ in Fig.~\ref{fig:2um_DCM_exp}~(b).  This correspondence allows us
to attribute the discontinuous feature at $\mu_0 H \approx -180$~mT in
$\Delta f (H)$, between (1) and (2), to the entrance of the first end
vortex. The entrance of the second vortex, marked by a square between
(2) and (3), though not visible in the measurement, produces a tiny
step in the simulated response at $\mu_0 H \approx -40$~mT.  Once at
$H = 0$, the FNT occupies configuration (3) with two end vortices and
an axially aligned central domain. Between $\mu_0 H \approx 10$ and
25~mT, an irreversible switching process causes the magnetization in
the central domain to flip, forming configuration (4) and producing a
change in the sign and slope of $\Delta f (H)$. The exit of the second
vortex between (4) and (5) can then be attributed to the discontinuity
at $\mu_0 H \approx 100$~mT, while the first vortex exits between (5)
and (6) producing the feature at $\mu_0 H \approx 200$~mT.

In Fig.~\ref{fig:2um_DCM_exp}~(c) we highlight the entrance and exit
of the first end vortex in both measured and simulated $\Delta f
(H)$. The hysteresis marks the first and last irreversible processes
of the magnetic reversal, indicating its nucleation and end,
respectively.  We find that both the magnitude in $\Delta f$ of the
simulated vortex entry and exit features and the field, at which they
occur, depend on the orientation of the FNT end surfaces (see
$\hat{n}_T$ and $\hat{n}_B$ in inset to Fig.~\ref{fig:setup}) with
respect to $\mathbf{H}$.  In the simulation, the angles of the ends
with respect to the FNT long axis, $\alpha_T$ and $\alpha_B$, are
carefully adjusted to match the measurements.  For the
2.2-\si{\micro\meter}-long FNT shown in Fig.~\ref{fig:2um_DCM_exp}
$\alpha_T \neq -\alpha_B$ ($\hat{n}_T \nparallel \hat{n}_B$),
resulting in two distinct pairs of entrance and exit fields. One
entrance and exit pair is barely visible in both experiment and
simulation due to that end's specific orientation with respect to
$\mathbf{H}$~\cite{SuppMat}.  The strong negative spike in the
measurements is not reproduced by the simulations. Although the origin
of this feature is unclear, changes in $\Delta f$ toward more negative
values correspond to a reduction in the angular magnetic confinement,
indicating a disordered intermediate magnetic configuration.
Simulations predicting DCM signatures of vortex entrance and exit in
FNTs have been carried out before, however, no corresponding features
were measured, likely due to the lack of well-defined ends, e.g.\
jagged ends or ends terminated by a growth-induced spherical
shell~\cite{gross_dynamic_2016}.

\begin{figure}
\includegraphics[]{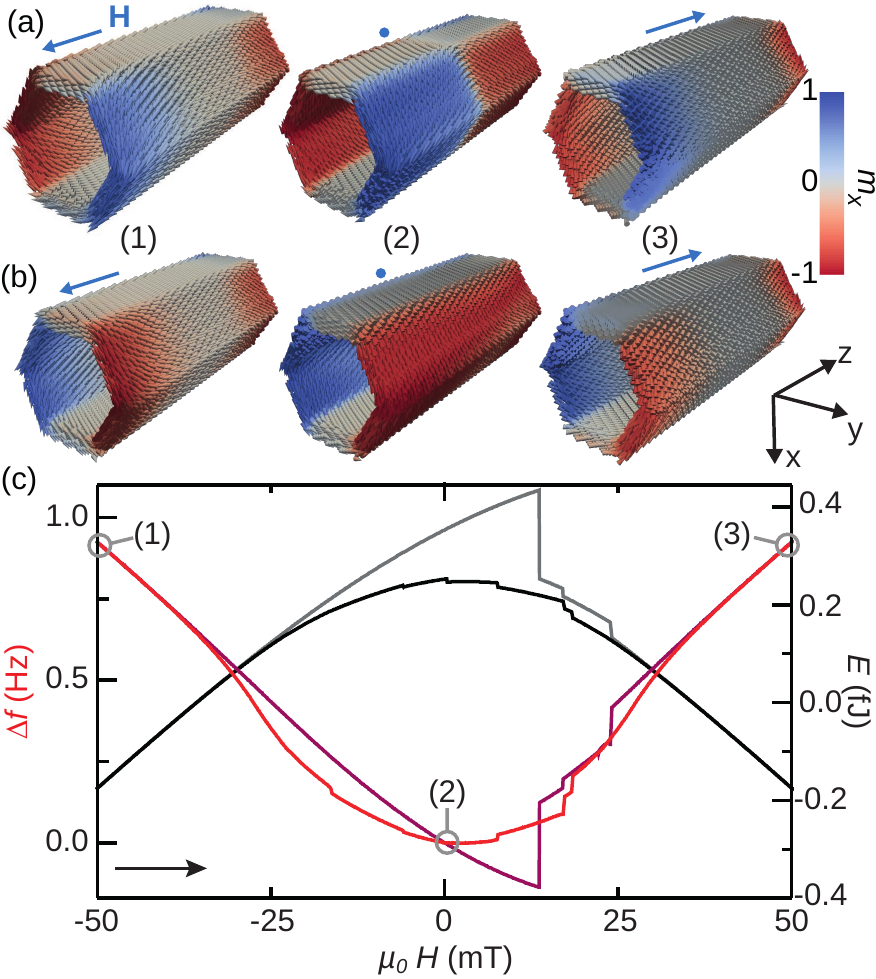}
\caption{Simulated reversal of a 0.6-\si{\micro\meter}-long FNT.  Equilibrium
  magnetization configurations corresponding to the labeled points in
  $\Delta f (H)$ for a FNT initialized with vortex ends of (a)
  opposing and (b) matching circulation sense.  (c) Plots of the
  simulated $\Delta f (H)$ in purple (red) and $E_m(H)$ in gray
  (black) for vortices of opposing (matching) circulation sense.  For
  the simulation, $\alpha_T = 6.0^\circ$, $\alpha_B = 10.0^\circ$, and
  $\theta_H = 10.0^\circ$.}
\label{fig:640nm_sim_opposing}
\end{figure}
Although the overall features of the measured and simulated $\Delta f
(H)$ for the 2.2-\si{\micro\meter}-long FNT match, there is a
difference in the irreversible switching of the central domain (around
$\mu_0 H \approx \pm 20$~mT in Fig.~\ref{fig:2um_DCM_exp}~(b)). The
measured response shows two distinct steps interrupted by a
plateau-like feature, rather than the single step predicted by the
simulations.  Measurements at slightly different $\theta_H$ and of the
2.9-\si{\micro\meter}-long FNT result in one to three such plateaus in
the switching region~\cite{SuppMat}.  These features indicate the
presence of intermediate magnetization configurations near zero
field~\cite{SuppMat}.

For short FNTs -- FNTs less than 2-\si{\micro\meter}-long for our cross-sectional
dimensions -- a different reversal process emerges.  Since during
reversal the two end vortices extend far enough to meet at the center
of the FNT, the two relative circulation senses of the end domains
lead to two different progressions, shown in
Fig.~\ref{fig:640nm_sim_opposing}~(a) and (b).  For end domains of
opposing circulation sense, simulations show that after the entrance
of the vortices, the central axial domain shrinks until only a domain
wall remains to separate the two vortex
domains~\cite{chen_equilibrium_2007,chen_magnetization_2011}.  As $H$
becomes increasingly positive, the axial wall reverses in a series of
irreversible steps, which are associated with the replacement of the
axial wall in each facet with a Bloch or Neel-type vortex wall. After
full reversal of the axial wall, the vortex domains recede, and
exit.

\begin{figure}
  \includegraphics[]{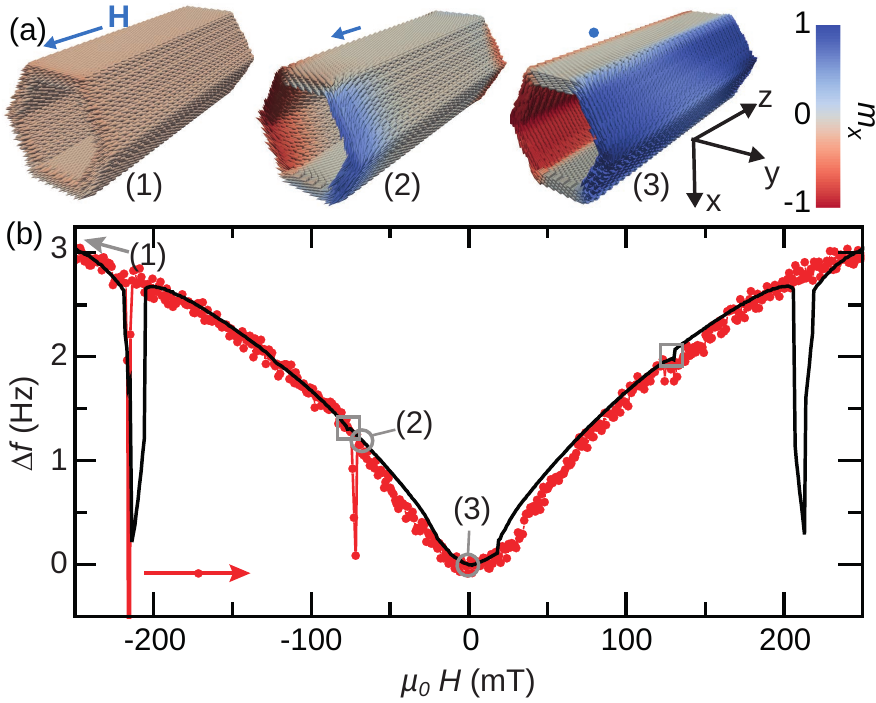}
  \caption{Simulated and measured reversal of a 0.6-\si{\micro\meter}-long FNT.
    (a) Calculated magnetization configurations for $\Delta f(H)$
    corresponding to the labels.  (b) Simulated (lines) and measured
    (points) DCM response.  Squares highlight simulated vortex
    entry/exit features, which are difficult to see.  For the
    simulation, $\alpha_T = 4.0^\circ$, $\alpha_B = 6.5^\circ$, and
    $\theta_H = 10.0^\circ$.}
\label{fig:640nm_exp_sim}
\end{figure}
For end domains of matching circulation sense, simulations show a
progression, in which the two vortex domains merge at the center of
the FNT without forming a domain wall. In reverse field, this global
vortex configuration progressively rotates toward $\mathbf{H}$, until
it splits and the resulting end vortices exit as the FNT
saturates. Steps occurring during this rotation are associated with
the switching of the magnetization in each hexagonal edge, where two
facets meet.

In Fig.~\ref{fig:640nm_sim_opposing}~(c), we plot simulations of both
$\Delta f (H)$ and the magnetic energy $E_m (H)$ associated with these
two reversal progressions.  Although end vortices with equal
circulation represent the lower energy remanent configuration in short
thin FNTs~\cite{chen_magnetization_2010,wyss_imaging_2017}, both
configurations have the same energy at high field within the accuracy
of our simulation.  Given the energy cost of switching between
matching and opposing configurations, FNTs should -- in principle --
reverse via both reversal progressions. In fact, experiments on
similar FNTs find both configurations in remanence after the
application of an axial field~\cite{wyss_imaging_2017}, confirming the
possibility of both reversal processes.  Note that simulations plotted
in Fig.~\ref{fig:640nm_sim_opposing}~(c) predict distinct $\Delta f
(H)$ signatures for short FNTs with end vortices of different relative
circulation sense.

As shown in Fig.~\ref{fig:640nm_exp_sim}, the measured DCM response of
the 0.6-\si{\micro\meter}-long FNT matches the progression with
vortices of matching circulation sense.  $\Delta f (H)$ never drops
below zero and is parabolic around zero field, indicating the presence
of a remanent global vortex state.  Features corresponding to the
entrance and exit of the two end vortices match in field magnitude and
to some extent also in $\Delta f$, with the exception of the large
feature connected with the exit of the first vortex.  This discrepancy
is likely due to fine details of the end geometry not captured by our
model.

\begin{figure}
\includegraphics[]{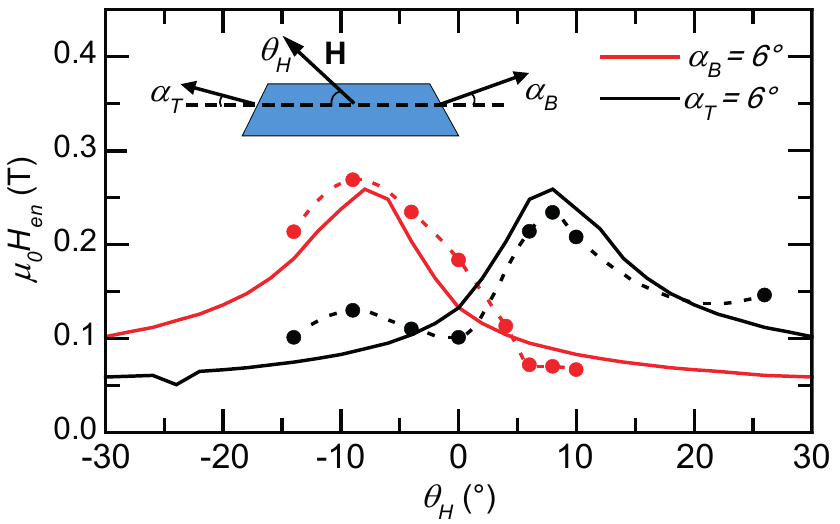}
\caption{Simulated and measured dependence of vortex entrance field on
  field angle.  Black (red) points show the measured $H_{en}$ as a
  function of $\theta_H$ for the top (bottom) end vortex in a single
  FNT.  Black (red) solid lines show the corresponding simulations for
  $\alpha_T = 6.0^\circ$ ($\alpha_B = 6.0^\circ$).  The schematic
  diagram depicts the FNT, its angled ends, and $\mathbf{H}$.}
\label{fig:640nm_entrance_field}
\end{figure}
In all simulations, we tune the orientation of the plane in which the
FNT ends lie ($\hat{n}_T$ and $\hat{n}_B$) with respect to
$\mathbf{H}$ in order to reproduce the measured features in $\Delta f
(H)$ associated with vortex entry and exit. We study this dependence
in more detail by measuring DCM in the 0.6-\si{\micro\meter}-long FNT
as a function of $\theta_H$.  Fig.~\ref{fig:640nm_entrance_field}
shows the experimentally determined and simulated entrance fields
$H_{en}$ of the top (bottom) vortex domain in black (red) as a
function of $\theta_H$.  The corresponding exit fields, which are not
shown, vary analogously.  Measurements and simulations show that
$H_{en}$ exhibits its absolute maximum just past $\theta_H = \pm
\alpha_{T/B}$, i.e.\ $\mathbf{H} \parallel \hat{n}_{T/B}$. Upon a
slight tilt of $\mathbf{H}$ away from this condition, $H_{en}$ is
reduced.  We attribute this behavior to the avoidance of magnetic
surface charge density $\sigma$.  In a saturated FNT, $\sigma$ at the
ends is maximized for $\mathbf{H} \parallel \hat{n}_{T/B}$.  As a
consequence, this alignment also maximizes the reduction in
magnetostatic energy resulting from the entrance of a vortex.  At
$H_{en}$, this reduction is equal to the corresponding Zeeman and
exchange energy penalties.  Given that the Zeeman penalty scales with
field, a maximum in $H_{en}$ is expected for $\mathbf{H} \parallel
\hat{n}_{T/B}$.  The close agreement between experiment and simulation
in Fig.~\ref{fig:640nm_entrance_field} suggests that the simulated
reversal nucleation process is an accurate description of the process
occurring in the measured samples.  Further simulations confirm that,
for a fixed orientation of $\mathbf{H}$, $H_{en}$ -- and therefore the
reversal nucleation field -- can be reduced from 250 to under 25~mT by
increasing the the slant angle $\alpha_{T/B}$ from 0 to
30$^\circ$~\cite{SuppMat}.

In conclusion, we find that even slightly slanted ends considerably
shift the nucleation field for axial magnetization reversal in FNTs.
Still, the magnetization reversal process is observed to occur through
vortex configurations, as originally predicted.  This experimental
confirmation of vortex-nucleated reversal and the demonstrated
tunability of the vortex entry field set the stage for the realization
of FNTs with fast and highly reproducible switching behavior.

\acknowledgments We thank S. Martin and his team in the machine shop of the Physics
Department at the University of Basel for help building the
measurement system.  We acknowledge the support of Kanton Aargau, the
Swiss Nanoscience Institute, the SNF under Grant No. 200020-159893,
the NCCR Quantum Science and Technology (QSIT), and the DFG via
project GR1640/5-2 in SPP 153.

\appendix

\section{Ferromagnetic Nanotube Fabrication} The template nanowires
(NWs), onto which the CoFeB shell forming the ferromagnetic nanotubes
(FNTs) is sputtered, are grown by molecular beam epitaxy on a Si (111)
substrate using Ga droplets as catalysts
\cite{ruffer_anisotropic_2014}.  During CoFeB deposition, a wafer of
upright and well-separated GaAs NWs is mounted with a 35$^\circ$ angle
between the long axis of the NWs and the deposition direction.  The
wafer is then continuously rotated in order to achieve a conformal
coating.  We cut individual FNTs into segments of different lengths
and with well-defined ends using a focused ion beam (FIB).  After
cutting, we use an optical microscope equipped with precision
micro-manipulators to pick up each FNT segment and affix it to the end
of an ultrasoft Si cantilever.  Non-magnetic epoxy~(Gatan G1) is used
as an adhesive.

\begin{figure}[h]
\includegraphics{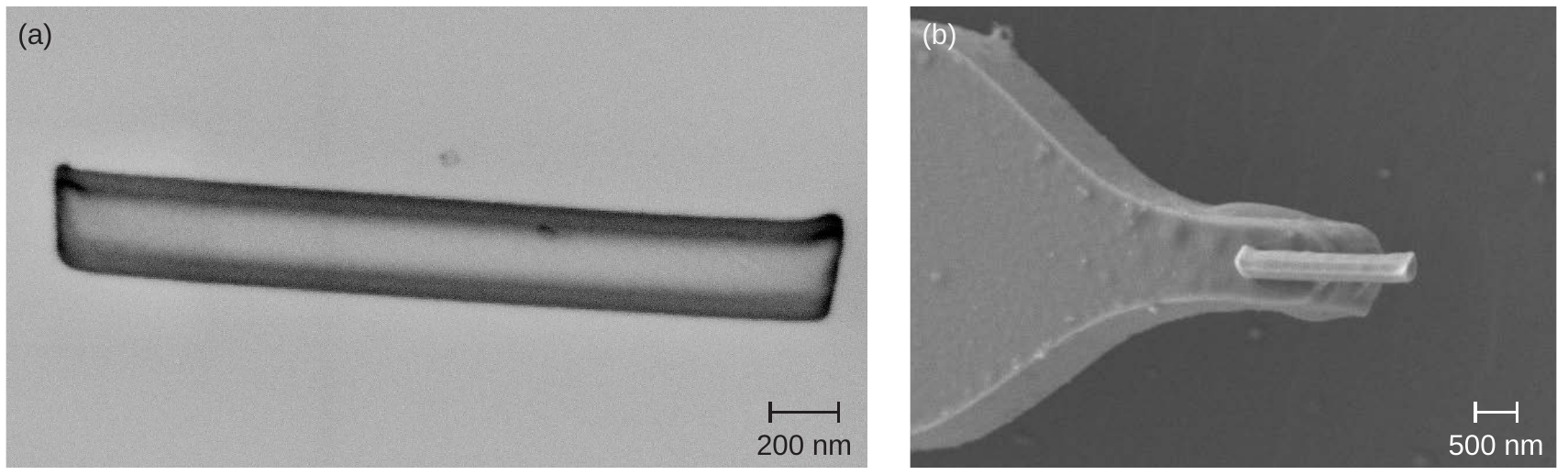}
\caption{Scanning electron micrographs (SEMs) of the FIB milled
  2.2~$\mu$m-long FNT (a) placed on a Si-surface and (b) attached to
  the tip of a ultra-soft Si cantilever.}
\label{fig:SEM2um}
\end{figure}

\begin{figure}[h]
\includegraphics{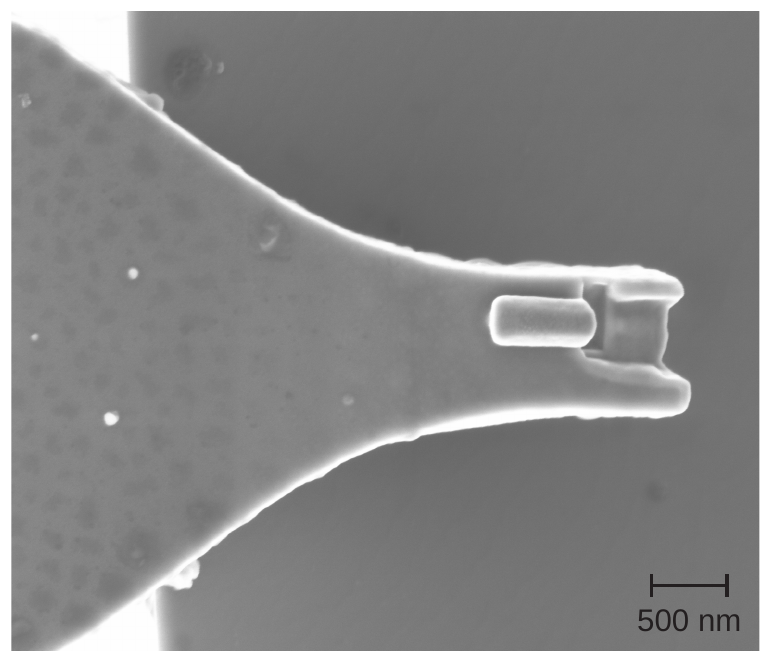}
\caption{SEMs of the 0.6-$\mu$m-long FNT attached to the tip of a
  Si-cantilever. The FNT was shortened in a second FIB step after the
  FNT was already attached to the cantilever.}
\label{fig:SEM680nm}
\end{figure}

\section{Dynamic Cantilever Magnetometry} The dynamic cantilever
magnetometry (DCM) measurement setup consists of a vibration-isolated
vacuum chamber with a pressure below $10^{-4}$~mbar.  A separate
manually rotatable superconducting magnet allows the application of an
external magnetic field $\mathbf{H}$ up to 4~T in any direction in the
$xz$-plane.  We use cantilevers made of undoped single-crystal Si with
a length of 150~$\mu$m, width of 4~$\mu$m, and thickness of
0.1~$\mu$m.  The spring constant $k_0 = 90$~$\mu$N/m and the effective
length of the fundamental mode $l_e = 105$~$\mu$m.  Unlike the others,
the cantilever used with the 2.2~$\mu$m-long FNT is 180~$\mu$m long,
$k_0 = 70$~$\mu$N/m and $l_e = 130$~$\mu$m.  Deflection of the
cantilever along $\hat{x}$ is measured using a fiber
interferometer~\cite{rugar_improved_1989} with 100~nW of 1550~nm laser
light focused onto a 10-$\mu$m-wide paddle near the end of the
cantilever.  A piezo-electric actuator mechanically drives the
cantilever at its resonance frequency with a constant oscillation
amplitude of 40~nm using a feed-back loop implemented by a
field-programmable gate array.  This process of self-oscillation
enables the fast and accurate extraction of the resonance frequency
$f$ from the cantilever deflection signal.  Before measurement, we
stabilize the temperature and fully magnetize the sample at large
$\mathbf{H}$.  DCM data is then collected as the field is stepped
toward zero and into reverse field.

\section{Mumax3 Simulations} We set $\mu_0 M_S$ to its measured value
of $\SI{1.3}{\tesla}$~\cite{gross_dynamic_2016} and the exchange
stiffness to $A_{ex} = \SI{28}{\pico\joule}/\si{\meter}$. We model the
FNTs as perfectly hexagonal tubes with slanted ends. Discretization of
space with cubic mesh elements leads to a staircase effect on all
slanted surfaces, which could have a impact on the magnetic states
that are calculated to be stable.  In order to exclude spurious
results due to such simulation artifacts, we perform reference
simulations with the finite element package \textit{nmag}
\cite{fischbacher_systematic_2007}, which avoids staircase effects by
using irregular tetragonal meshes.  In particular, \textit{nmag}
simulations reveal the same stable magnetization configurations, the
same vortex entry mechanism, and the same values for the vortex entry
(exit) field $H_{en}$ ($H_{ex}$).  As a result, we conclude that the
staircase effect on the FNT ends does not have a significant effect on
our simulation results.

Both \textit{Mumax3} and \textit{nmag} are used to determine the
equilibrium magnetization configuration for each external field value
by numerically solving the Landau-Lifshitz-Gilbert equation.  Since
the microscopic processes in FNTs are expected to be much faster than
the cantilever resonance frequency \cite{landeros_domain_2010,
  schwarze_magnonic_2013, yan_chiral_2012, gross_dynamic_2016}, the
magnetization of the nanotube can always be assumed to be in its
equilibrium orientation.  The calculation also yields the total
magnetic energy $E_m$ corresponding to each configuration.  In order
to simulate $\Delta f$ measured in DCM, we numerically calculate the
second derivative of $E_m$ with respect to $\theta_c$ found in (1) in
the main text.  At each field, we calculate $E_m$ at the cantilever
equilibrium angle $\theta_c = 0$ and at small deviations from
equilibrium $\theta_c = \pm \delta \theta_c$.  For small $\delta
\theta_c$, the second derivative can by approximated by a finite
difference: $\left.  \frac{\partial^2 E_m}{\partial \theta_c^2} \right
|_{\theta_c=0} \approx \frac{E_m(\delta \theta_c) - 2 E_m(0) +
  E_m(-\delta \theta_c)}{(\delta \theta_c)^2}$. By setting $f_0$,
$k_0$, and $l_e$ to their measured values, we then arrive at the
$\Delta f$ corresponding to each magnetization configuration in the
numerically calculated field dependence.  Table~\ref{tab:sim_params}
shows the exact parameters for the geometry of the FNT and the
direction of the applied field used for the simulations shown in the
main text.

\begin{table}[h]
  \caption{Parameters used for the simulations shown in the main text.  $d$ is the FNT diameter, $t$ its thickness, $l$ its length, $\alpha_T$ ($\alpha_B$) the slant angle of its top (bottom) end, $\theta_H$ ($\phi_H$) is the polar (azimuthal) angle of  $\mathbf{H}$, $\delta \theta_c$ is the cantilever deviation angle used to calculate $\Delta f$, and $\epsilon_m$ is the mesh size. }
\label{tab:sim_params}
\begin{ruledtabular}
  \begin{tabular}{lccccccccc} Fig. & $d$ (nm) & $t$ (nm) & $l$ ($\mu$m) & $\alpha_T$ ($^\circ$) & $\alpha_B$ ($^\circ$) & $\theta_H$ ($^\circ$) & $\phi_H$ ($^\circ$) & $\delta \theta_c$ ($^\circ$) & $\epsilon_m$ (nm)\\
    \colrule
    3 & 280 & 30 & 2.180 & 6.5 & 10.5 & 11.0 & -6.0 & 2.0 & 5 \\
    4 & 284 & 30 & 0.640 & 6.0 & 10.0 & 10.0 & 1.0 & 1.0 & 4 \\
    5 & 284 & 30 & 0.640 & 4.0 & 6.5 & 10.0 & 1.0 & 1.0 & 4 \\
    6 & 284 & 30 & 0.640 & 6.0 & 6.0 & - & 1.0 & 1.0 & 4 \\
    \end{tabular}
\end{ruledtabular}
\end{table}

\section{Reversal Measured at Low Temperature}

\begin{figure}[h]
\includegraphics{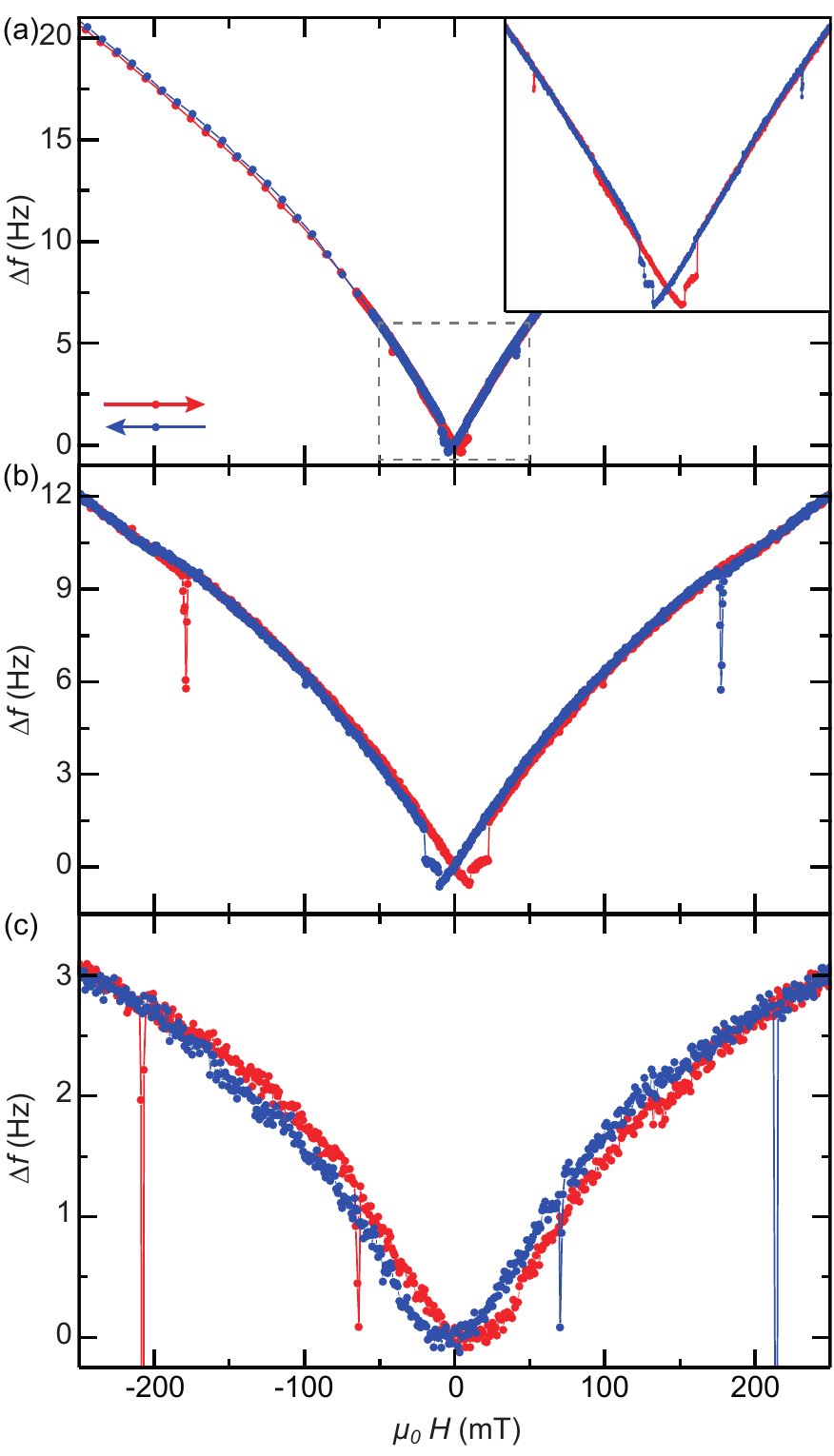}
\caption{Magnetic reversal of the three FNTs of different lengths
  measured by DCM at 280~K: (a) the 2.9-$\mu$m-long, including a zoom of
  the low field region; (b) the 2.2-$\mu$m-long; and (c) the
  0.6-$\mu$m-long FNT.}
\label{fig:df280K}
\end{figure}

\begin{figure}[h]
\includegraphics{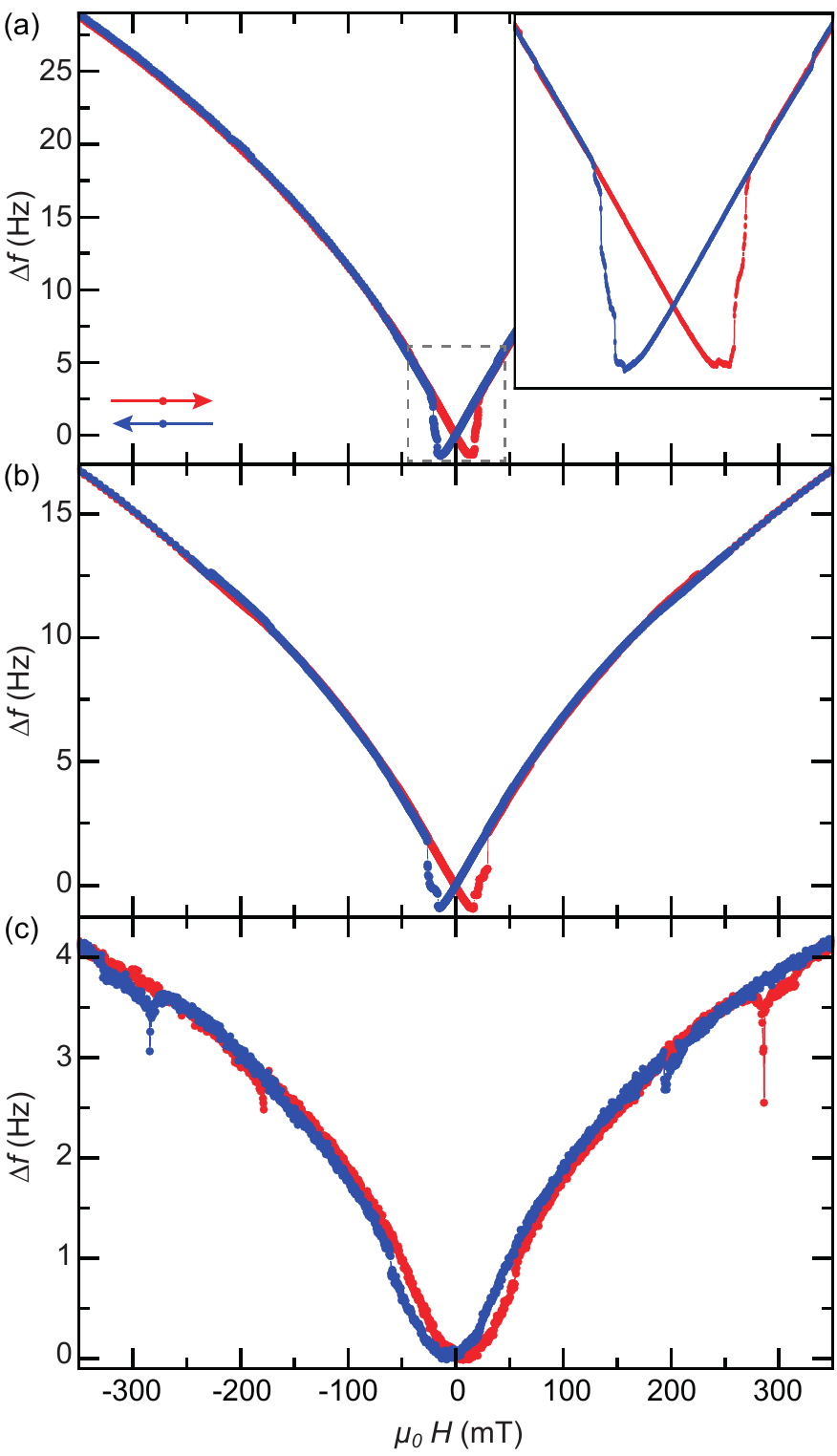}
\caption{Magnetic reversal of the three FNTs of different lengths
  measured by DCM at 4~K: (a) the 2.9-$\mu$m-long, including a zoom of
  the low field region; (b) the 2.2-$\mu$m-long; and (c) the
  0.6-$\mu$m-long FNT.}
\label{fig:df5K}
\end{figure}

Fig.~\ref{fig:df280K} shows DCM measurements at 280~K of three FNTs of
different lengths.  Measurements of a 2.9-$\mu$m-long FNT are shown in
(a), while measurements of a 2.2-$\mu$m-long FNT and 0.6-$\mu$m-long
FNT, which already appear in the main text, are shown in (b) and (c),
respectively.  As expected from numerical simulations and theory, the
FNTs longer than 2 $\mu$m display a qualitatively similar behavior
corresponding to a common magnetization reversal process.

Fig.~\ref{fig:df5K} shows a second set of DCM measurements of the same
three FNTs carried out at 4~K.  Although the same qualitative features
observed at 280~K can be recognized, various details of the reversal
differ.  First, the features in $\Delta f(H)$ indicating the entrance
or exit of a vortex are less pronounced at low temperature than at
280~K (shown in Fig.~2).  Second, for the two longer FNTs, the
hysteric region marking an irreversible switching process spans a
larger field range at low temperature than at high temperature.  This
behavior reflects the smaller amount of thermal energy available to
the system at low temperature to overcome the energy barriers impeding
magnetization reversal.  Although the entrance and exit of the two
vortices appear at similar fields for all FNTs at both 4~K and 280~K,
differences in the angle of the magnetic field $\theta_H$ for each
measurement preclude drawing conclusions about the dependence of
entrance/exit field on temperature.

\section{Plateau in $\Delta f$ for small applied fields }

\begin{figure}[b]
\includegraphics{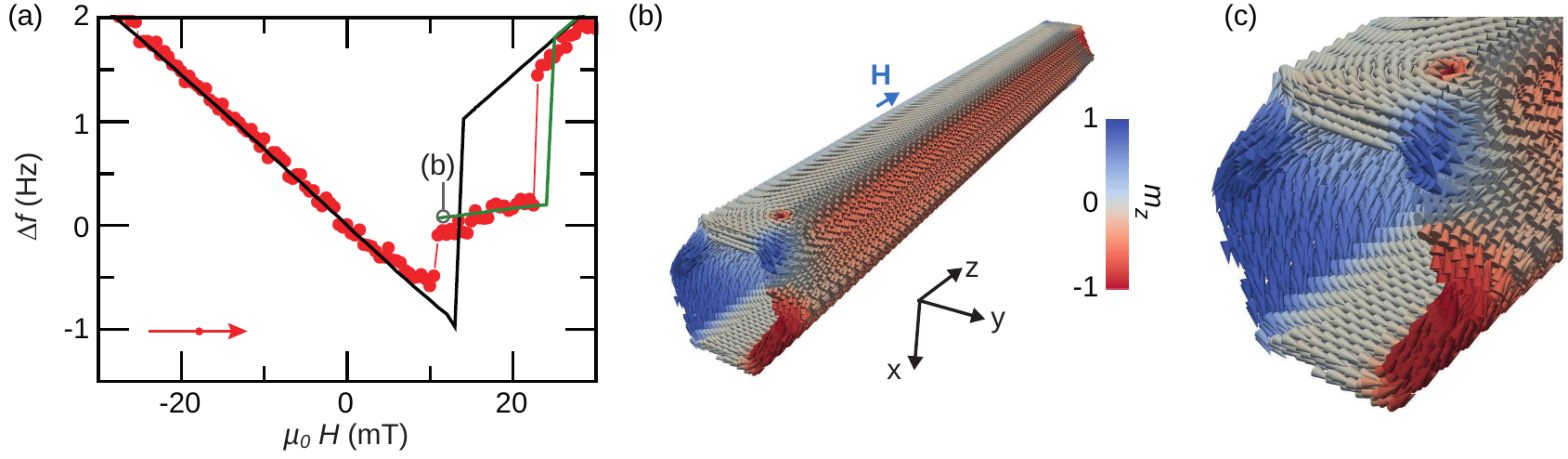}
\caption{(a) A detailed view of the simulated (line) and measured
  (points) DCM signatures of the irreversible switching at low-field
  for the 2.2-$\mu$m-long FNT.  The green line shows the DCM response
  of a magnetization configuration, shown in (b) and (c), which is
  initialized and calculated to be stable at $H =0$.  The
  configuration includes two vortices each residing in a hexagonal
  facet of the FNT and produces a $\Delta f (H)$, which matches the
  plateau appearing in the measured response.}
\label{fig:FacetVortex}
\end{figure}

DCM measurements shown in Fig.~2~(a) and Fig.~\ref{fig:df5K}~(a) and
(b) show plateau-like features in the irreversible switching region
around zero field that are not predicted by the simulations.  The
behavior in $\Delta f (H)$ can be reproduced, however, by initializing
the FNT configuration at $H = 0$.  For example, by initializing the
FNT with two vortices each residing in a hexagonal facet of the FNT
and sweeping $H$ from zero to positive fields, a plateau in the
simulated $\Delta f (H)$ emerges, as shown in
Fig.~\ref{fig:FacetVortex}~(a).  This plateau feature corresponds to
an intermediate configuration with two facet vortices, shown in
Fig.~\ref{fig:FacetVortex}~(b) and (c).  Following the irreversible
switch around 25~mT, these facet vortices 'rotate' around $\hat{y}$
and take their place as the end vortices of an FNT in a mixed state
configuration.  In this picture, the irreversible switching of the
FNTs central axial domain is characterized by the 'rotation' of end
vortices into the facets -- consuming the axial domain and resulting
in the plateau in $\Delta f (H)$ -- followed by a second rotation of
the vortices from the facets back to the ends.  Similar states are
present in the simulations for short FNTs with opposing vortex
circulation sense.  For example, the plateau between 12 and 25~mT in
the purple $\Delta f (H)$-curve in Fig.~4~(c) is the result of a
configuration with two facet vortices.  Although this reversal mode is
a possibility which matches the measured $\Delta f (H)$ near zero
field, we cannot rule out the possibility of other intermediate
configurations resulting in the same DCM response.

\section{Simulated and Measured Vortex Entrance}

\begin{figure}[h]
  \includegraphics{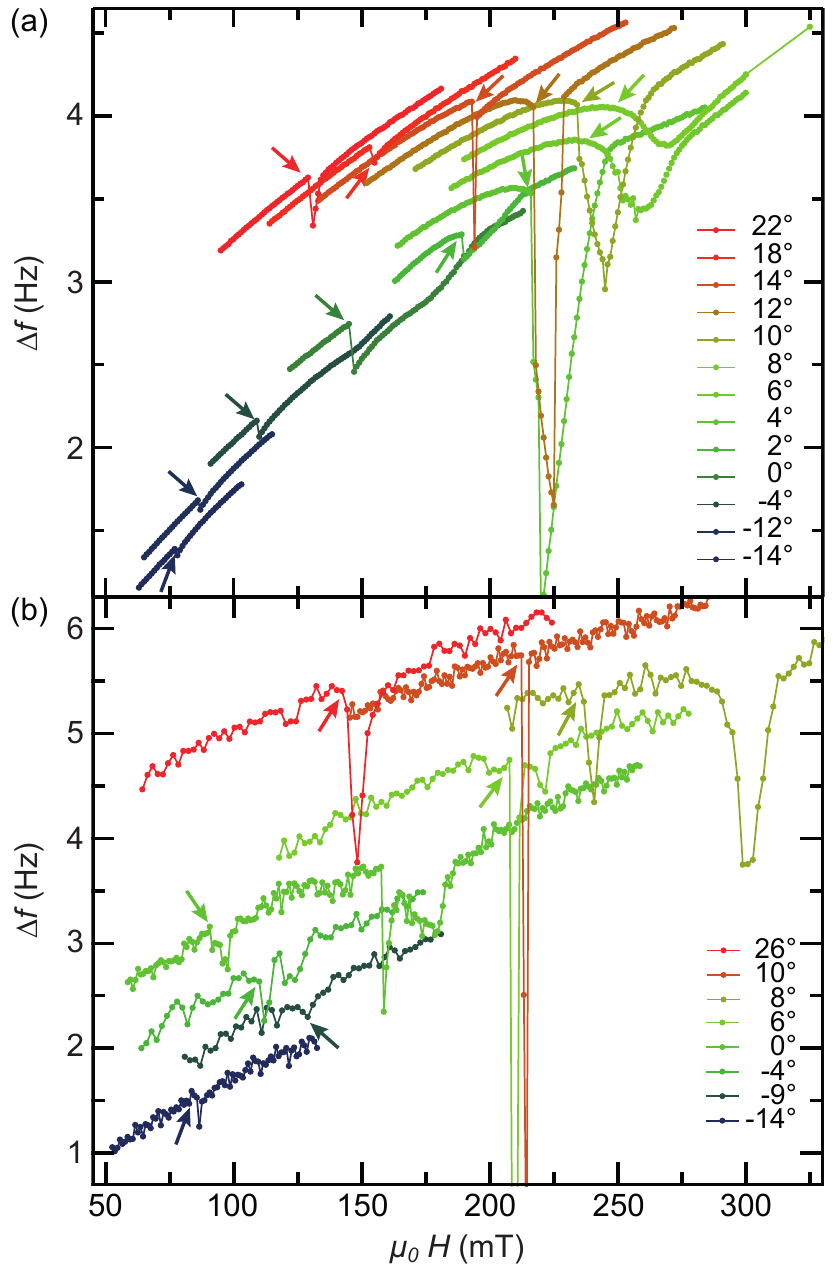}
  \caption{(a) simulated and (b) measured segments of $\Delta f (H)$
    showing the vortex formation in a 0.6~$\mu$m-long NT for different
    values of $\theta_H$ as labeled in the plots.  Arrows highlight
    the specific feature corresponding to the vortex entrance, as
    indicated in corresponding simulations of the FNT magnetization
    configuration.}
\label{fig:vortexEntrance}
\end{figure}

Fig.~6 of the main text illustrates the dependence of the vortex
entrance on the magnetic field angle $\theta_H$.  The data plotted in
that figure are extracted from the DCM measurements shown in
Fig.~\ref{fig:vortexEntrance}.  This figure shows a comparison between
(a) simulated and (b) measured $\Delta f (H)$ in the region of one
vortex entrance for the 0.6-$\mu$m-long FNT.  Note the strong
dependence of the magnitude in $\Delta f$ of the entrance features on
$\theta_H$.  As with the value of $H_{en}$, this magnitude is
maximized for $\mathbf{H} \parallel \hat{n}_{T/B}$.  In other words,
the vortex entrance and exit features for an end whose normal is
strongly misaligned with $\mathbf{H}$ are nearly invisible by DCM.

\section{Control of Reversal Nucleation Field}

\begin{figure}[h]
\includegraphics{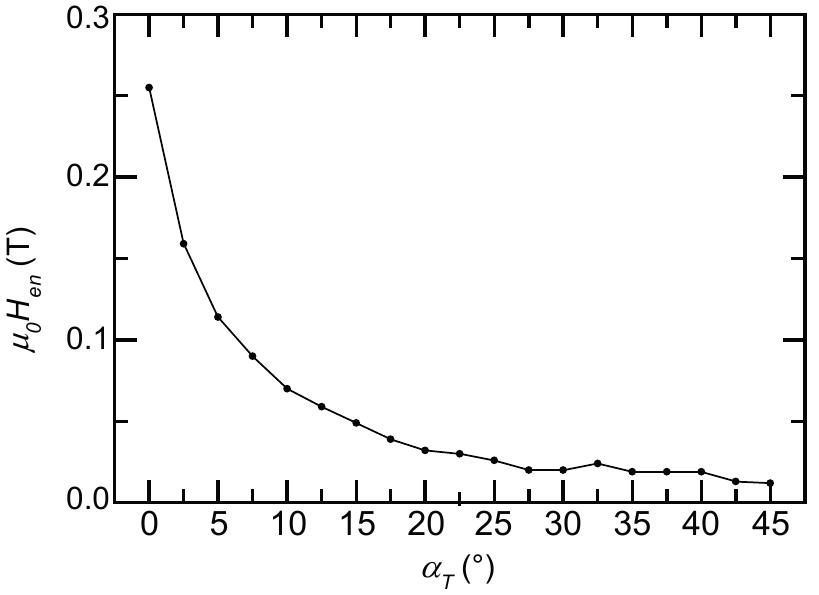}
\caption{Simulated dependence of the entrance field $H_{en}$ of the
  top vortex on the slant angle of the top end $\alpha_T$ for the
  0.6-$\mu$m-long FNT.  The magnetic field is applied parallel to
  $\hat{z}$, i.e.\ $\theta_H = 0$.}
\label{fig:vortexSlanting}
\end{figure}

Fig.~6 makes clear that the angle of the applied magnetic field
$\theta_H$ affects the entrance field of the vortex and thus the
reversal nucleation field of the FNT.  Fig.~\ref{fig:vortexSlanting}
shows the simulated dependence of the entrance of the top vortex on
the slant angle of the top FNT end $\alpha_T$ for a fixed $\theta_H$.
The entrance field can be tuned by over 225~mT by changing the slant
angle by $30^\circ$.  These simulations, combined with the
experimental evidence shown in the main text, show that reversal
nucleation in FNTs can be finely and predictably controlled by tuning
the geometry of their ends.

\section{Simulation of Vortex Entrance}

The supplementary animation
\href{https://youtu.be/wkKhsbd9-yo}{VortexEntrance} shows the
progression of magnetization configurations present in the
0.6-$\mu$m-long FNT as the magnetic field is reduced from positive
values towards zero.  At the same time, the animation shows the
simulated values of $\Delta f (H)$, making clear the correspondence
between a discontinuous feature in $\Delta f (H)$ -- similar to those
observed in our measurements -- and the vortex entrance.  In this
particular simulation, the field is applied with an angle $\theta_H =
14^\circ$ with respect to the long axis of the FNT.

\end{document}